\documentstyle[epsfig,12pt]{article}
\newcommand{\la}[1]{\label{#1}}
\newcommand{\ur}[1]{(\ref{#1})}
\newcommand{\urs}[2]{(\ref{#1},\ref{#2})}

\newcommand{\eq}[1]{eq.~(\ref{#1})}
\newcommand{\eqs}[2]{eqs.~(\ref{#1},\ref{#2})}
\newcommand{\Eq}[1]{Eq.~(\ref{#1})}
\newcommand{\Eqs}[2]{Eqs.(\ref{#1}, \ref{#2})}
\newcommand{\eqss}[3]{eqs.~(\ref{#1},\ref{#2},\ref{#3})}
\newcommand{\n}{\bf{n}}

\def\Tr{\mbox{Tr}}
\def\beq{\begin{equation}}
\def\eeq{\end{equation}}
\def\bea{\begin{eqnarray}}
\def\eea{\end{eqnarray}}

\begin{document}
\thispagestyle{empty}
\begin{flushright} NORDITA-2000/65 HE
\end{flushright}
\vskip 2true cm
\begin{center}
{\Large\bf Non-Abelian Stokes Theorems in Yang--Mills \\
\vskip .2true cm
and Gravity Theories}

\vskip 1.5true cm

{\large\bf Dmitri Diakonov$^{\diamond *}$ and Victor Petrov$^*$} \\
\vskip 1true cm
$^\diamond$ {\it NORDITA, Blegdamsvej 17, DK-2100 Copenhagen \O,
Denmark} \\
\vskip .5true cm
$^*$ {\it St.Petersburg Nuclear Physics Institute, Gatchina 188 350,
Russia} \\
\vskip .5true cm

E-mail: diakonov@nordita.dk, victorp@thd.pnpi.spb.ru
\end{center}
\vskip 1.5true cm
\begin{abstract}
\noindent We recall the non-Abelian Stokes theorem for the
Wilson loop in the Yang --Mills theory and discuss its meaning.
Then we move to `gravitational Wilson loops', i.e. to holonomies in
curved $d=2,3,4$ spaces and derive `non-Abelian Stokes theorems' for
these quantities as well, which are similar to our formula in the
Yang--Mills theory. In particular we derive an elegant formula for the
holonomy in the case of a constant-curvature background in three
dimensions and a formula for small-area loops in any number of dimensions.
\end{abstract}

\section{Introduction}

One of the main objects in the Yang--Mills theory as well as in
gravity is the parallel transporter along closed contours, or holonomy.
In Yang--Mills theory it is conventionally called the Wilson loop; it can
be written as a path-ordered exponent,

\beq
W_r=
\frac{1}{d(r)}\Tr \;{\rm P}\;
\exp\, i\oint\!d\tau\,\frac{dx^\mu}{d\tau}\,A_\mu^a\,T^a,
\la{wl1}\eeq
where $x^\mu(\tau)$ with $0\leq\tau\leq 1$ parametrizes the closed
contour, $A_\mu^a$ is the Yang--Mills field (or connection) and $T^a$
are the generators of the gauge group in a given representation $r$
whose dimension is $d(r)$. In curved Riemannian spaces the
`gravitational Wilson loop' or holonomy for $d$-dimensional vectors
can be also written as a trace of the path-ordered exponent of the
connection, this time of the Christoffel symbol,

\beq
W^G_{{\rm vector}}=
\frac{1}{d}\left[{\rm P}\;\exp\,
-\oint\!d\tau\,\frac{dx^\mu}{d\tau}\,\Gamma_\mu\right]^\kappa_\kappa.
\la{gwl1}\eeq
One can also consider parallel transporters of spinors in curved
background: in this case the holonomy is defined not by the Christoffel
symbols but by the spin connection which is not uniquely determined
by the metric tensor, see the precise definitions below.

The Yang--Mills Wilson loop is invariant under gauge transformations of
the background field $A_\mu$; the gravitational Wilson loop is
invariant under general coordinate transformations or diffeomorphisms,
provided one transforms the contour as well.

It is generally believed that in three and four dimensions the average
of the Wilson loop in a pure Yang--Mills quantum theory exhibits an
area behaviour for large and simple contours (like flat rectangular).
This should be true not for all representations but those
with `$N$-ality' nonequal zero; in the simplest case of the $SU(2)$
gauge group these are representations with half-integer spin $J$.

One of the difficulties in proving the asymptotic area law for the
Wilson loop in half-integer representations (and proving that in
integer representations it is absent) is that the Wilson
loop is a complicated object by itself: it is impossible to calculate it
analytically in a general non-Abelian background field. Meanwhile,
it is sometimes easier to average a quantity over an ensemble than to
calculate it for a specific representative. However, in case of the
Wilson loop the path-ordering is a serious obstacle on that way.

A decade ago we have suggested a formula for the Wilson
loop in a given background belonging to any gauge group and any
representation \cite{DP1}. In this formula the path ordering along the
loop is removed, but at the price of an additional integration over all
gauge transformations of the given non-Abelian background field,
or, more precisely, over a coset depending on the particular
representation in which the Wilson loop is considered. Furthermore, the
Wilson loop can be presented in a form of a surface integral \cite{DP2},
see the next section. We call this representation the non-Abelian Stokes
theorem. It is quite different from previous interesting statements
\cite{Halp,Ar,Br,Sim} also called by their authors `non-Abelian Stokes
theorem' but which involve surface ordering. Our formula has no surface
ordering. A classification of `non-Abelian Stokes theorems' for
arbitrary groups and their representations has been given recently by
Kondo et al. \cite{Kondo} who used the naturally arising techniques of
flag manifolds.

Though these formulae usually do not facilitate finding Wilson
loops in particular backgrounds, they can be used to average Wilson
loops over ensembles of Yang--Mills configurations or over different
metrics, and in more general settings, see. e.g.
\cite{DP3,Pol1,Kondo,KAK}.

The main aim of this paper is to present new formulae for the
gravitational holonomies in curved $d=2,3,4$ spaces: they are
similar to our non-Abelian Stokes theorem for the Yang--Mills case. We
get rid of the path ordering in \eq{gwl1} and write down the holonomies
as exponents of surface integrals. Instead of path ordering we have
to integrate over certain covariantly unit vectors (in case of $d=3$)
or covariantly unit (anti)self-dual tensors (in case of $d=4$).
Remarkably, these formulae put parallel transporters of different spins
on the same footing. In particular, holonomies for half-integer
spins are presented in terms of the metric tensor (and its derivatives)
only but not in terms of the vielbein or spin connection.

Apart from purely theoretical interest we have a practical motivation
in mind. Quite recently we have shown, both in the continuum and on the
lattice, that the $SU(2)$ Yang--Mills partition function in $d=3$ can
be exactly rewritten in terms of local gauge-invariant quantities
being the six components of the metric tensor of the dual space
\cite{DP4}. This rewriting may be useful to investigate the spectrum
and the correlation functions of the theory directly in a
gauge-invariant way, but it is insufficient to study the interactions of
external sources since they couple to the Yang--Mills potential and not
to gauge-invariant quantities. The present paper demonstrates, however,
that a typical source, i.e. the Yang-Mills Wilson loop can be expressed
not only through the potential (or connection) but also through the
metric tensor which is gauge-invariant. Thus, not only the partition
function but also Wilson loops in the $d=3$ Yang--Mills theory can be
expressed through local gauge-invariant quantities. We leave a detailed
formulation of the resulting theory for a forthcoming publication.

Though the main content of the paper are the non-Abelian Stokes
theorems for holonomies in 3 and 4 dimensions, we have added three
short sections with relevant material. We add for completeness
the Stokes theorem in two dimensions, compute the holonomy in a special
case of constant curvature with cylinder topology in three dimensions
and give a general formula for the `gravitational Wilson loop' for
small loops in any number of dimensions.

\section{Non-Abelian Stokes theorem in Yang--Mills theory}

Let $\tau$ parametrize the loop defined by the trajectory
$x^\mu(\tau)$ and $A(\tau)$ be the tangent component of the Yang--Mills
field along the loop in the fundamental representation of the gauge
group, $A(\tau)=A_\mu^at^adx^\mu/d\tau$, $\mbox{Tr}(t^at^b) =
\frac{1}{2}\delta^{ab}$. The gauge transformation of $A(\tau)$ is

\beq
A(\tau) \rightarrow S(\tau)A(\tau)S^{-1}(\tau)
+iS(\tau)\frac{d}{d\tau}S^{-1}(\tau).
\la{gt}\eeq
Let $H_i$ be the generators from the Cartan subalgebra ($i=1,...,r;\;
r$ is the rank of the gauge group) and the $r$-imensional vector
${\bf m}$ be the highest weight of the representation $r$ in which
the Wilson loop is considered. The formula for the Wilson loop derived
in ref. \cite{DP1} is a path integral over all gauge transformations
$S(\tau)$ which should be periodic along the contour:

\beq
W_r=\int DS(\tau) \exp\,i\int d\tau\; \Tr\,\left[m_iH_i\;
(SAS^{-1} +i S\dot S^{-1})\right].
\la{W1}\eeq
Let us stress that \eq{W1} is manifestly gauge invariant, as is the
Wilson loop itself.  For example, in the simple case of the $SU(2)$
group \eq{W1} reads:

\beq
W_J=\int DS(\tau) \exp\,i\,J\int d\tau \;\Tr\,\left[\tau_3
(SAS^\dagger +i S\dot S^\dagger)\right]
\la{W12}\eeq
where $\tau_3$ is the third Pauli matrix and
$J=\frac{1}{2},\;1,\; \frac{3}{2},...$
is the `spin' of the representation of the Wilson loop considered.

The path integrals over all gauge rotations \urs{W1}{W12} are not
of the Feynman type: they do not contain terms quadratic in the
derivatives in $\tau$. Therefore, a certain regularization is
understood in these equations, ensuring that $S(\tau)$ is sufficiently
smooth. For example, one can introduce quadratic terms in the angular
velocities $iS\dot S^\dagger$ with small coefficients eventually put to
zero; see ref.\cite{DP1} for details. In ref.\cite{DP1}  \eq{W12} has
been derived in two independent ways: i) by direct discretization and
ii) by using the standard Feynman representation of path integrals as a
sum over all intermediate states, in this case that of an axial top
supplemented by a `Wess--Zumino' type of the action. Another
discretization but leading to the same result has been used recently by
Kondo \cite{Kondo}. A similar formula has been used by Alekseev,
Faddeev and Shatashvili \cite{AFS} who derived a formula for group
characters to which the Wilson loop is reduced in case of a constant
$A$ field actually considered in \cite{AFS}. In ref.\cite{Lun} \eq{W1}
has been rederived in an independent way specifically for the
fundamental representation of the $SU(N)$ gauge group. Finally, another
derivation of a variant of \eq{W12} using lattice regularization has
been presented recently in ref. \cite{DPnew}.

The second term in the exponent of \eqs{W1}{W12} is in fact a
`Wess--Zumino'-type action, and it can be rewritten not as
a line but as a surface integral inside a closed contour.
Let us consider for simplicity the $SU(2)$
gauge group and parametrize the $SU(2)$ matrix $S$ from \eq{W12} by
Euler's angles,
\bea
\nonumber
S&=&\exp(i\gamma \tau_3/2)\;\exp(i\beta\tau_2/2)\;
\exp(i\alpha\tau_3/2)\\
\nonumber
\\
\la{Euler}
&=&\left(\begin{array}{cc}
\cos\frac{\beta}{2}\,e^{i\frac{\alpha+\gamma}{2}}&
\sin\frac{\beta}{2}\,e^{-i\frac{\alpha-\gamma}{2}}\\
-\sin\frac{\beta}{2}\,e^{i\frac{\alpha-\gamma}{2}}&
\cos\frac{\beta}{2}\,e^{-i\frac{\alpha+\gamma}{2}}
\end{array}\right).
\eea
The derivation of \eq{W12} implies that $S(\tau)$ is a periodic matrix.
It means that $\alpha\pm\gamma$ and $\beta$ are periodic functions of
$\tau$, modulo $4\pi$.

The second term in the exponent of \eq{W12} which we denote by $\Phi$
is then
\bea
\nonumber
\Phi &=& \int d\tau\;\Tr(\tau_3iS\dot S^\dagger)
=\int d\tau\, \dot \alpha (\cos\beta+\dot \gamma)\\
\la{WZ1}
&=& \int d\tau\, [\dot \alpha (\cos\beta-1)+(\dot \alpha+\dot \gamma)]
=\int d\tau\,\dot \alpha (\cos\beta-1).
\eea
The last term is a full derivative and can be actually dropped because
$\alpha+\gamma$ is
$4\pi$-periodic, therefore even for half-integer representations $J$
it does not contribute to \eq{W12}. Notice that $\alpha$ can be
$2\pi$-periodic if $\gamma$ (which drops from \eq{WZ1}) is
$2\pi,\,6\pi,\ldots$-periodic. If $\alpha(1)=\alpha(0)+2\pi\,k$,
$\alpha(\tau)$ makes $k$ windings. Integration over all possible
$\alpha(\tau)$ implied in \eq{W12} can be divided into distinct sectors
with different winding numbers $k$.

Introducing a unit 3-vector

\beq
n^a=\frac{1}{2} \Tr\;(S\tau^a S^\dagger \tau_3)
=(\sin\beta\cos\alpha, \,\sin\beta\sin\alpha, \,\cos\beta)
\la{n}\eeq
we can rewrite $\Phi$ as

\beq
\Phi=\frac{1}{2}\int d\tau d\sigma\,\epsilon^{abc}\,\epsilon_{ij}
\,n^a\partial_i n^b\partial_j n^c,\qquad i,j=\tau,\sigma,
\la{WZ2}\eeq
where one integrates over any surface spanned on the contour (we shall
call it a `disk'), and ${\bf n}$ or $\alpha$ and $\beta$ are continued
inside the disk without singularities. We denote the second coordinate by
$\sigma$ such that $\sigma=1$ corresponds to the edge of the disk
coinciding with the contour and $\sigma=0$ corresponds to the center of
the disk. See ref. \cite{DPnew} for details on the continuation inside
the disk.

Let us note that if the surface is closed or infinite the r.h.s. of
\eq{WZ2} is the integer topological charge of the ${\bf n}$ field on
the surface:

\beq
Q=\frac{1}{8\pi}\int d\sigma d\tau\,\epsilon^{abc}\,\epsilon_{ij}
\,n^a\partial_i n^b\partial_j n^c.
\la{topc}\eeq

\Eq{WZ2} can be also rewritten in a form which is invariant under
the reparametrizations of the surface. Introducing the invariant
element of a surface,

\beq
d^2S^{\mu\nu}=d\sigma\,d\tau\;\left(
\frac{\partial x^\mu}{\partial \tau}
\frac{\partial x^\nu}{\partial \sigma}-
\frac{\partial x^\nu}{\partial \tau}
\frac{\partial x^\mu}{\partial \sigma}\right)
= \epsilon^{\mu\nu}\;d({\rm Area}),
\la{elsur}\eeq
one can rewrite \eq{WZ2} as

\beq
\Phi=\frac{1}{2}\int\; d^2S^{\mu\nu}
\epsilon^{abc} n^a\partial_\mu n^b\partial_\nu n^c.
\la{WZ3}\eeq
We get for the Wilson loop \cite{DP1}:

\beq
W_J=\int D{\bf n}(\sigma,\tau)\;\exp\left[iJ\int\,d\tau
(A^an^a)+\frac{iJ}{2}\int\;d^2S^{\mu\nu}
\epsilon^{abc} n^a\partial_\mu n^b\partial_\nu n^c\right].
\la{W2}\eeq

The interpretation of this formula is obvious: the unit vector $\n$ plays
the role of the instant direction of the colour `spin' in colour space;
however, multiplying its length by $J$ does not yet guarantee that we
deal with a true quantum state from a representation labelled by $J$ --
that is achieved only by introducing the `Wess--Zumino' term in \eq{W2}:
it fixes the representation to which the probe quark of the Wilson loop
belongs to be exactly $J$.

Finally, we can rewrite the exponent in \eq{W2} so that both terms
appear to be surface integrals \cite{DP2}:

\beq
W=\int D{\bf n}(\sigma,\tau)\;\exp\frac{iJ}{2}\int
d^2S^{\mu\nu}\left(-F_{\mu\nu}^an^a+
\epsilon^{abc}\, n^a\left(D_\mu n\right)^b\left(D_\nu
n\right)^c\right),
\la{W3}\eeq
where
$D_\mu^{ab}=\partial_\mu\delta^{ab}+\epsilon^{acb}A_\mu^c$
is the covariant derivative and $F_{\mu\nu}^a = \partial_\mu~
A_\nu^a - \partial_\nu~A_\mu^a + \epsilon^{abc}~A_\mu^b~A_\nu^c$
is the field strength. Indeed, expanding the exponent of \eq{W3} in
powers of $A_\mu$ one observes that the quadratic term cancels out
while the linear one is a full derivative reproducing the $A^an^a$ term
in \eq{W2}; the zero-order term is the `Wess--Zumino' term \ur{WZ2} or
\ur{WZ1}. Note that both terms in \eq{W3} are explicitly gauge
invariant. We call \eq{W3} the non-abelian Stokes theorem. We stress
that it is different from previously suggested Stokes-like
representations of the Wilson loop, based on ordering of elementary
surfaces inside the loop \cite{Halp,Ar,Br,Sim}. For a further discussion
of \eq{W3} see \cite{DPnew}.

Let us briefly discuss gauge groups higher than $SU(2)$: for that
purpose we have to return to our \eq{W1}. \Eq{W1} is valid
for any group and any representation. It is easy to present it also in a
surface form, see ref. \cite{DPnew}. Actually, \eq{W1} depends not on
all parameters of the gauge transformation but only on those which do
not commute with the Cartan combination ${\cal H}_r=m_iH_i$. In the
$SU(2)$ case one has $m_iH_i= J\tau_3,\;J=1/2,1,3/2,\ldots$, since
$SU(2)$ is of rank 1, and there is only one Cartan generator.
Therefore, in the $SU(2)$ case one integrates over the coset
$SU(2)/U(1)$ for any representation; this coset can be parametrized by
the ${\bf n}$ field as described above.

In case of higher groups the particular coset depends on the
representation of the Wilson loop. For example, in case the Wilson loop
is considered in the fundamental representation of the $SU(N)$ group the
combination $m_iH_i$ is proportional to one particular generator of the
Cartan subalgebra, which commutes with the $SU(N-1)\times U(1)$ subgroup.
[In case of $SU(3)$ this generator is the Gell-Mann $\lambda_8$
matrix or a permutation of its elements.] Therefore, the appropriate
coset for the fundamental representation of the $SU(N)$ group is
$SU(N)~/SU(N-1)~/U(1)= CP^{N-1}$. A possible parametrization of this
coset is given by a complex $N$-vector $u^\alpha$ of unit length,
$u_\alpha^\dagger u^\alpha=1$. To be concrete, the Cartan combination
in the fundamental representation can be always set to be
$m_iH_i={\rm diag}(1,0,\ldots,0)$ by rotating the axes and
subtracting the unit matrix. In such a basis $u^\alpha$ is just the
first column of the unitary matrix $S$ while $u_\alpha^\dagger$
is the first row of $S^\dagger$. Unitarity of $S$ implies that
$u_\alpha^\dagger u^\alpha=1$.

In this parametrization \eq{W1} can be written as

\beq
W^{SU(N)}_{{\rm fund}}=\int Du\,Du^\dagger\,
\delta(u_\alpha^\dagger u^\alpha-1)\,
\exp i\!\int\! d\tau\,\frac{dx^\mu}{d\tau}
u^\dagger_\alpha\left(i\nabla_\mu\right)^\alpha_\beta\,u^\beta,
\qquad \left(\nabla_\mu\right)^\alpha_\beta
=\partial_\mu\,\delta^\alpha_\beta-iA^a_\mu
\left(t^a\right)^\alpha_\beta.
\la{W13}\eeq
Using the identity,
\bea
\nonumber
\epsilon_{ij}\,\partial_i\left(u^\dagger\nabla_iu\right)&=&
\epsilon_{ij}\,\left[\left(\nabla_i u\right)^\dagger
\left(\nabla_j u\right)+u^\dagger\nabla_i\nabla_ju\right]\\
\la{stokes}
&=&\epsilon_{ij}\,\left[-\frac{i}{2}(u^\dagger F_{ij}u)+
\left(\nabla_i u\right)^\dagger\left(\nabla_j u\right)\right],
\eea
we can present \eq{W13} in a surface form:

\beq
W^{SU(N)}_{{\rm fund}}=\int Du\,Du^\dagger\,
\delta(|u|^2-1)\,
\exp i\!\int\! dS^{\mu\nu}
\left[\frac{1}{2}(u^\dagger F_{\mu\nu}u)+i
\left(\nabla_\mu u\right)^\dagger\left(\nabla_\nu u\right)\right],
\la{W14}\eeq
where $F_{\mu\nu}$ is the field strength in the fundamental
representation. \Eq{W14} has been first published in ref.\cite{Lun}
however with an unexpected overall coefficient 2 in the exponent.
\Eq{W14} presents the non-Abelian Stokes theorem for the Wilson loop
in the fundamental representation of $SU(N)$. In the particular case
of the $SU(2)$ group transition to \eq{W3} is achieved by identifying
the unit 3-vector: $n^a=u^\dagger_\alpha(\tau^a)^\alpha_\beta u^\beta$
where

\beq
u^\alpha=\left(\begin{array}{c}\cos\frac{\beta}{2}\,
e^{-i\frac{\alpha+\gamma}{2}}\\
\sin\frac{\beta}{2}\,
e^{\,i\frac{\alpha-\gamma}{2}}\end{array}\right),\qquad
2i\,u^\dagger\partial_\tau u
=\dot\alpha(\cos\beta-1)+(\dot\alpha+\dot\gamma).
\la{u}\eeq

It should be mentioned that the quantity

\beq
\int d\sigma d\tau \epsilon_{ij}\,i\partial_iu^\dagger_\alpha
\partial_ju^\alpha= 2\pi Q
\la{topzar}\eeq
appearing in \eq{W14} is the topological charge of the 2-dimensional
$CP^{N-1}$ model. For closed or infinite surfaces $Q$ is an integer.

In case the Wilson loop is taken in the adjoint represention of
the $SU(N)$ gauge group the combination $m_iH_i$ is the highest root.
Only group elements of the form $\exp(i\alpha_iH_i)$ commute with this
combination, belonging to the maximal torus subgroup $U(1)^{N-1}$.
Hence, in case of the adjoint representation one in fact integrates
over the maximal coset $SU(N)/~U(1)^{N-1}=F^{N-1}$, i.e. over flag
variables \cite{Per,Kondo}.

\section{`Gravitational Wilson loops'}

An object similar to the Wilson loop of the Yang--Mills theory exists
also in gravity theory. It is the parallel transporter of a vector
on a Riemannian manifold along a closed contour, else called a
holonomy. The holonomy is trivial if the space is flat but becomes
a non-trivial functional of the curvature in case it is nonzero.
In the remaining sections we shall present new formulae for the parallel
transporters on $d=2,3,4$ Riemannian manifolds.

Let us first remind notations from differential geometry. We
use \cite{LogL} as a general reference book.

Let $g_{\mu\nu}=g_{\nu\mu}\;(\mu,\nu=1,\ldots d)$ be the covariant
metric tensor, with the contravariant $g^{\mu\nu}$ being its inverse,
$g_{\mu\nu}g^{\nu\kappa}=\delta^\kappa_\mu$. The determinant of the
covariant metric tensor is denoted by $g$. The Christoffel symbol is
defined as

\beq
\Gamma^\mu_{\nu\kappa}=g^{\mu\lambda}\Gamma_{\lambda,\nu\kappa}
=\frac{g^{\mu\lambda}}{2}(\partial_\nu g_{\lambda\kappa}
+\partial_\kappa\,g_{\lambda\nu}-\partial_\lambda\,g_{\nu\kappa}),
\qquad\Gamma^\kappa_{\nu\kappa}=
\frac{\partial_\nu\,g}{2g}.
\la{Christ}\eeq
The action of the covariant derivative on the contravariant
vector is defined as

\beq
(\nabla_\rho)^\kappa_\lambda v^\lambda
=(\partial_\rho\delta^\kappa_\lambda
+\Gamma^\kappa_{\rho\lambda})v^\lambda.
\la{covder}\eeq
The commutator of two covariant derivatives determine the Riemann
tensor:

\beq
\left[\nabla_\rho\nabla_\sigma\right]^\kappa_\lambda
=R^\kappa_{\;\lambda\rho\sigma}=g^{\kappa\kappa^\prime}
R_{\kappa^\prime \lambda\rho\sigma}=
\partial_\rho\Gamma^\kappa_{\sigma\lambda}
-\partial_\sigma\Gamma^\kappa_{\rho\lambda}
+\Gamma^\kappa_{\rho\tau}\Gamma^\tau_{\sigma\lambda}
-\Gamma^\kappa_{\sigma\tau}\Gamma^\tau_{\rho\lambda}.
\la{Riem1}\eeq
A contraction of the Riemann tensor gives the symmetric Ricci tensor,

\beq
R_{\lambda\sigma}=R^\kappa_{\;\lambda\kappa\sigma},
\qquad R^\kappa_\rho=R^\kappa_{\;\lambda\rho\sigma}g^{\lambda\sigma}.
\la{Ricci1}\eeq
Its full contraction is the scalar curvature:

\beq
R=R_{\lambda\sigma}g^{\lambda\sigma}=R^\kappa_\kappa .
\la{sccurv1}\eeq

The parallel transporter of a contravariant vector along a curve
$x^\mu(\tau)$ is determined by solving the equation,
\beq
\frac{dx^\mu}{d\tau}\,(\nabla_\mu)^\kappa_\lambda\,v^\lambda(\tau)=0,
\la{partra1}\eeq
whose solution can be written with the help of the evolution operator,
\beq
v^\kappa(\tau)=\left[W^G(\tau)\right]^\kappa_\lambda\,v^\lambda(0)
\la{evop1}\eeq
where $v^\lambda(0)$ is the vector at the starting point of the contour
and $v^\lambda(\tau)$ is the parallel-transported vector at the point
labelled by $\tau$. The evolution operator can be symbolically written
as a path-ordered exponent of the Christoffel symbol:

\beq
\left[W^G(\tau)\right]^\kappa_\lambda
=\left[{\rm P}\;\exp\,-\int_0^\tau d\tau\frac{dx^\mu}{d\tau}
\Gamma_\mu\right]^\kappa_\lambda.
\la{wlgsymb}\eeq

We define the `gravitational Wilson loop' as the {\em trace} of the
parallel transporting evolution operator along the closed curve
$x^\mu(\tau)$ with $x^\mu(1)=x^\mu(0)$:

\beq
W^G_{{\rm vector}}=\frac{1}{d}\left[W^G(1)\right]^\kappa_\kappa.
\la{gravW1}\eeq
This quantity is diffeomorphism-invariant: if one changes the
coordinates $x^\mu\to x^{\prime\,\mu}(x)$ the metric is
transformed, but if one changes the contour accordingly, that is
$x^\mu(\tau)\to x^{\prime\,\mu}(x(\tau))$ the gravitational Wilson
loop or the holonomy remains the same. In this respect the
gravitational holonomy is different from the Yang--Mills Wilson
loop which is invariant under gauge transformation, without changing
the contour.

The parallel transporter of a covariant vector is given by the
transposed matrix; its trace coincides with that of the contravariant
vector.

\section{Relation of gravity quantities to those of the
Yang--Mills theory}

We shall show now that the `gravitational Wilson loop' is not just
analogous but directly expressible through the Yang--Mills Wilson loops
of the $SU(2)$ group. To that end we introduce the standard vielbein
$e^A_\mu$ and its inverse $e^{A\mu}$ such that

\beq
e^A_\mu e^A_\nu=g_{\mu\nu},\qquad e^A_\mu e^{B\mu}=\delta^{AB},\qquad
 e^{A\mu }e^{A\nu}=g^{\mu \nu},\qquad \det e^A_\mu  =\sqrt{g}.
\la{viel}\eeq
Let us decompose the vector experiencing the parallel transport in
vielbeins,

\beq
v^\lambda=c^A e^{A\lambda},\qquad {\rm the\; reciprocal\; being\;}\;\;
c^A =e^A_\kappa v^\kappa,
\la{decomp1}\eeq
and put it into \eq{partra1} defining the parallel transport.  We have

\beq
0=\frac{dx^\mu }{d\tau}(\nabla_\mu )^\kappa_\lambda c ^A e^{A\lambda}=
\frac{dx^\mu }{d\tau}\left[e^{A\kappa}\partial_\mu c^A
+c^A (\partial_\mu e^{A\kappa}+\Gamma^\kappa_{\mu\lambda}
e^{A\lambda})\right]
=\frac{dx^\mu }{d\tau}e^{B\kappa}
(\partial_\mu \delta^{BA}+\omega^{BA}_\mu )c^A ,
\la{partra2}\eeq
where we have introduced the spin connection,

\beq
\omega^{AB}_\mu =-\omega^{BA}_\mu
=\frac{1}{2}e^{A\kappa}(\partial_\mu e_\kappa^B-\partial_\kappa e_\mu^B)
-\frac{1}{2}e^{B\kappa}(\partial_\mu e_\kappa^A -\partial_\kappa e_\mu^A)
-\frac{1}{2}e^{A\kappa}e^{B\lambda}e^C_\mu
(\partial_\kappa e_\lambda^C-\partial_\lambda e_\kappa^C),
\la{conn}\eeq
and used the fundamental relation:

\beq
\partial_\mu
e^{A\kappa}+\Gamma^\kappa_{\mu\lambda}e^{A\lambda}
=-\omega^{A B}_\mu e^{B\kappa},
\la{e1}\eeq
\beq
\partial_\mu e^A_\kappa-\Gamma^\lambda_{\mu\kappa}e^A_\lambda
=-\omega^{AB}_\mu e^B_\kappa\,.
\la{e2}\eeq

One can introduce the $SO(d)$ `field strength',

\beq
{\cal F}^{AB}_{\mu\nu}=\left[\partial_\mu+\omega_\mu,
\partial_\nu+\omega_\nu\right]^{AB}=
\partial_\mu \omega^{AB}_\nu-\partial_\nu \omega^{AB}_\mu+
\omega^{AC}_\mu\omega^{CB}_\nu-\omega^{AC}_\nu\omega^{CB}_\mu,
\la{fsSO}\eeq
related to the Riemann tensor as

\beq
{\cal F}^{AB}_{\mu\nu}e^A_\kappa e^B_\lambda
=-R_{\kappa\lambda\mu\nu},\qquad
{\cal F}^{AB}_{\mu\nu}=-R_{\kappa\lambda\mu\nu}e^{A\kappa}e^{B\lambda},
\qquad {\cal F}^{AB}_{\mu\nu}e^{A\mu} e^{B\nu}=R.
\la{FtoR}\eeq

The above material is common for any number of dimensions. To proceed
further we need to consider separately the cases $d=3$ and $d=4$. The
case $d=2$ is considered in section 6.

\subsection{d=3}

In three dimensions one can immediately identify the spin connection
with a Yang--Mills field in $su(2)$,

\beq
A^c_i=-\frac{1}{2}\epsilon^{abc}\,\omega^{ab}_i.
\la{YMc}\eeq
Speaking about three dimensions we shall denote Lorentz indices by
$i,j,...=1,2,3$ and the flat triade indices by $a,b,...=1,2,3$.
Recalling the generators in the $J=1$ representation,

\beq
(T^c)^{ab}=-i\epsilon^{cab},\qquad [T^cT^d]=i\epsilon^{cdf}T^f,
\la{gens1}\eeq
we can rewrite the last parenthesis in \eq{partra2} as

\beq
\partial_i\delta^{ab}+\omega^{ab}_i =
\partial_i\delta^{ab}-iA^c_i(T^c)^{ab} \equiv (D_i)^{ab},
\la{covd1}\eeq
which is the standard Yang--Mills covariant derivative in the adjoint
representation. In the fundamental (spinor) representation the
Yang--Mills covariant derivative is

\beq
(\nabla_i)^\alpha_\beta=
\partial_i\delta^\alpha_\beta
-iA^c_i\left(\frac{\sigma^c}{2}\right)^\alpha_\beta
=\partial_i\delta^\alpha_\beta+\frac{1}{8}\omega^{ab}_i\,
\left[\sigma^a\sigma^b\right]^\alpha_\beta,
\qquad\alpha,\beta=1,2,
\la{covd2}\eeq
which coincides with the known expression for the covariant derivative
in the spinor representation in curved space.

The standard Yang--Mills field strength is directly related to
that of \eq{fsSO}:

\beq
F^a_{ij}=\partial_iA^a_j-\partial_jA^a_i+\epsilon^{abc}A^b_iA^c_j
=-\frac{1}{2}\epsilon^{abc}{\cal F}^{bc}_{ij}
\la{fs}\eeq
Hence from \eq{FtoR} one has

\beq
\epsilon^{abc}\,F^a_{ij}e^b_ke^c_l=R_{ijkl}.
\la{rel5}\eeq

Let us consider the parallel transporter of a 3-vector in curved space,
as defined by \eq{partra1}. According to \eqs{partra2}{covd1}
solving \eq{partra1} is equivalent to solving the Yang--Mills
equation for the parallel transporter,

\beq
\frac{dx^i}{d\tau}(D_i)^{ab}c^b=0,
\la{partra3}\eeq
whose solution is

\beq
c^a(\tau)=\left[W^{YM}_1(\tau)\right]^{ab}\,c^b(0),
\qquad \left[W^{YM}_1(\tau)\right]^{ab}
=\left[{\rm P}\, \exp \,
i\int\!d\tau\,\frac{dx^i}{d\tau}\,A_i^c\,T^c\right]^{ab}
\la{WYM}\eeq
where the subscript ``1'' refers to that the path-ordered exponent
is taken in the $J=1$ representation. The parallel transport of a
contravariant vector is therefore

\beq
v^k(\tau)=c^a(\tau)e^{ak}(\tau)=e^{ak}(\tau)
\left[W^{YM}_1(\tau)\right]^{ab}e^b_l(0)\,v^l(0),
\la{partra4}\eeq
from where we immediately get the needed relation between the
`gravitational' and Yang--Mills parallel transporters:

\beq
\left[W^G_1(\tau)\right]^k_l=e^a_k(\tau)\left[W^{YM}_1(\tau)\right]^{ab}
e^{bl}(0).
\la{rel1}\eeq
The relation becomes especially neat for the Wilson loops, i.e. for the
{\em traces} of parallel transporters along closed contours. Since for
the closed contour the vielbeins at the end points are the same,
$e^a_k(1)=e^a_k(0)$, we get

\beq
W^G_{{\rm vector}}=\frac{1}{3}\left[W^G_1\right]^k_k=
\frac{1}{3}\left[W^{YM}_1\right]^{aa}=W^{YM}_1.
\la{rel2}\eeq
In a similar way one can show that the same equation holds true for the
gravitational parallel transporter of {\em covariant} vectors
and, more generally, for parallel transporters of any integer spin $J$.
In this case the Yang--Mills Wilson loop should be taken in the
same representation as the gravitational one:

\beq
W^G_J=W^{YM}_J.
\la{rel3}\eeq

In \eq{rel3} it is understood that the r.h.s. is expressed through the
Yang--Mills field equal to the spin connection according to
\eq{YMc}, while the l.h.s. is expressed through the Christoffel
symbols, that is through the metric. It should be stressed that the
spin connection is defined via the vielbein which is not uniquely
determined by the metric tensor.  Nevertheless, the Wilson loop, being
a gauge-invariant quantity, is uniquely determined by the metric tensor
and its derivatives. This is the meaning of \eq{rel3}.

For half-integer $J$ there is no way to define the parallel transporter
other than through the spin connection. Nevertheless, as we show
in section 8 where we present the holonomy for any spin in a surface
form, the `gravitational Wilson loop' is also expressible through
the metric tensor and its derivatives, even for half-integer spins.

\subsection{d=4}

In four Euclidean dimensions the rotational group is $SO(4)=SU(2)\times
SU(2)$, therefore all irreducible representations of $SO(4)$ can be
classified as $(J_1,J_2)$ where $J_{1,2}=0,\frac{1}{2},1,...$ are the
representations of the two $SU(2)$ subgroups. For example, the 4-vector
representation whose parallel transporter has been considered in the
beginning of this section, transforms as the
$(\frac{1}{2},\frac{1}{2})$ representation of $SU(2)\times SU(2)$.

Because of this, it is convenient to decompose the spin connection
$\omega^{AB}_\mu$ into self-dual and anti-self-dual parts using
't Hooft's $\eta$ and $\bar\eta$ symbols. They are defined as
\bea
\la{eta1}
\eta^{aAB}&=&\frac{1}{2i}\Tr\,\sigma^a(\sigma^{A+}\sigma^{B-}-
\sigma^{B+}\sigma^{A-}),\qquad \sigma^{A\pm}=(\pm i{\bf \sigma},1),\\
\la{etb1}
\bar\eta^{aAB}&=&\frac{1}{2i}\Tr\,\sigma^a(\sigma^{A-}\sigma^{B+}-
\sigma^{B-}\sigma^{A+}).
\eea
We use capital Latin characters to denote flat 4-dimensional
vierbein indices, $A,B,\ldots=1,2,3,4$, while $a,b,\ldots=1,2,3$;
$\sigma^a$ are the three Pauli matrices. The spin connection
$\omega^{AB}_\mu$ transforms as a 6-dimensional representation
of $SO(4)$, which can be decomposed as a sum $(1,0)+(0,1)$ of
adjoint representations of the two $SU(2)$ subgroups. We write:

\beq
\omega^{AB}_\mu=-\frac{1}{2}\,\pi^a_\mu\,\eta^{aAB}
-\frac{1}{2}\,\rho^a_\mu\,\bar\eta^{aAB}.
\la{omdec}\eeq
The $SO(4)$ `field strength' \ur{fsSO} is then decomposed accordingly:

\beq
{\cal F}^{AB}_{\mu\nu}=-\frac{1}{2}\,F^a_{\mu\nu}(\pi)\,\eta^{aAB}
-\frac{1}{2}\,F^a_{\mu\nu}(\rho)\,\bar\eta^{aAB}
\la{fsSOdec}\eeq
where
\bea
\la{fspi}
F^a_{\mu\nu}(\pi)&=&\partial_\mu\pi^a_\nu-\partial_\nu\pi^a_\mu
+\epsilon^{abc}\pi^b_\mu\pi^c_\nu,\\
\la{fsrho}
F^a_{\mu\nu}(\rho)&=&\partial_\mu\rho^a_\nu-\partial_\nu\rho^a_\mu
+\epsilon^{abc}\rho^b_\mu\rho^c_\nu
\eea
are the usual Yang--Mills field strengths of the
$SU(2)$ Yang--Mills potentials $\pi^a_\mu$ and $\rho^a_\mu$.
It should be stressed that $6\cdot 4=24$ variables $\omega^{AB}_\mu$
equivalent to $2\cdot 3\cdot 4=24$ variables $\pi^a_\mu,\,\rho^a_\mu$
are defined by only $4\cdot 4=16$ tetrades $e^A_\mu$ via \eq{conn}, so
that not all of them are independent.

Contracting \eq{FtoR} with the $\eta,\bar\eta$ symbols we get:
\bea
\la{FpitoR}
F^a_{\mu\nu}(\pi)&=&\frac{1}{2}\eta^{aAB}e^{A\kappa}e^{B\lambda}
R_{\kappa\lambda\mu\nu},\\
\la{FrhotoR}
F^a_{\mu\nu}(\rho)&=&\frac{1}{2}\bar\eta^{aAB}e^{A\kappa}e^{B\lambda}
R_{\kappa\lambda\mu\nu}.
\eea

Let us now return to the parallel transporter of a 4-vector. As shown
in the beginning of this section, finding it is equivalent to solving
the equation

\beq
\frac{dx^\mu}{d\tau}\left(\partial_\mu\,\delta^{AB}+
\omega^{AB}_\mu\right)\,c^B=0.
\la{partra5}\eeq
Let us present the 4-vector $c^A$ as a combination of two spinors,

\beq
c^A=\chi^\dagger_\alpha\left(\sigma^{A+}\right)^\alpha_\beta\psi^\beta,
\qquad\chi^\dagger_\alpha\psi^\beta=\frac{1}{2}c^A
\left(\sigma^{A-}\right)^\beta_\alpha,
\qquad\alpha,\beta=1,2.
\la{4v}\eeq
Putting it into \eq{partra5} and decomposing $\omega^{AB}_\mu$ as
in \eq{omdec} we get:

\beq
\frac{dx^\mu}{d\tau}\left\{
\partial_\mu\left[\chi^\dagger\sigma^{A+}\psi\right]
-\frac{1}{2}\left(\pi^a_\mu\,\eta^{aAB}+\rho^a_\mu\,\bar\eta^{aAB}\right)\,
\left[\chi^\dagger\sigma^{B+}\psi\right]\right\} =0.
\la{partra6}\eeq
Using the definition of the $\eta$-symbols \urs{eta1}{etb1} it is easy
to check that this equation is satisfied provided the spinors
$\chi,\,\psi$ satisfy
\bea
\la{covpi}
\frac{dx^\mu}{d\tau}\left[\partial_\mu\,\delta^\alpha_\beta
-i\,\pi^a_\mu\left(\frac{\sigma^a}{2}\right)^\alpha_\beta\right]
\chi^\beta&=&0\quad{\rm or}\quad
\frac{dx^\mu}{d\tau}\,\chi^\dagger_\alpha
\left[\overleftarrow\partial_\mu\,\delta^\alpha_\beta
+i\,\pi^a_\mu\left(\frac{\sigma^a}{2}\right)^\alpha_\beta\right]
=0,\\
\la{covrho}
\frac{dx^\mu}{d\tau}\left[\partial_\mu\,\delta^\alpha_\beta
-i\,\rho^a_\mu\left(\frac{\sigma^a}{2}\right)^\alpha_\beta\right]
\psi^\beta&=&0.
\eea
Expressions in square brackets are identical to the Yang--Mills
covariant derivatives, the role of the Yang--Mills potentials played
by $\pi^a_\mu$ and $\rho^a_\mu$, respectively.
\Eqs{covpi}{covrho} define the Yang--Mills parallel transporters
in the fundamental representation. Their solution can be written as
evolution operators,
\bea
\la{partrapi}
\chi^\alpha(\tau)&=&\left[W^\pi(\tau)\right]^\alpha_\gamma\chi^\gamma(0)
\quad{\rm or}\quad
\chi^\dagger_\alpha(\tau)=\chi^\dagger_\gamma(0)
\left[W^{\pi\dagger}(\tau)\right]^\gamma_\alpha,\\
\la{partrarho}
\psi^\beta(\tau)&=&\left[W^\rho(\tau)\right]^\beta_\delta\psi^\delta(0), \\
\la{Wpi}
\left[W^\pi(\tau)\right]^\alpha_\gamma
&=&\left[{\rm P}\, \exp \,
i\int\!d\tau\,\frac{dx^\mu}{d\tau}\,\pi_\mu^a\,\frac{\sigma^a}{2}
\right]^\alpha_\gamma,\\
\la{Wrho}
\left[W^\rho(\tau)\right]^\alpha_\gamma
&=&\left[{\rm P}\, \exp \,
i\int\!d\tau\,\frac{dx^\mu}{d\tau}\,\rho_\mu^a\,\frac{\sigma^a}{2}
\right]^\alpha_\gamma.
\eea
Returning to the 4-vector $c^A$ \ur{4v} we see that its evolution is
determined by
\bea
\nonumber
c^A(\tau)&=&\left[W_{{\rm vector}}(\tau)\right]^{AB}\,c^B(0),\\
\la{opev4v}
\left[W_{{\rm vector}}(\tau)\right]^{AB}&=&
\frac{1}{2}\,\Tr\left[W^{\pi\dagger}(\tau)\sigma^{A+}W^\rho(\tau)
\sigma^{B-}\right].
\eea
Let us take a closed contour and take the trace of the evolution
operator. The `gravitational Wilson loop' for a 4-vector is then
\bea
\nonumber
W^G_{(\frac{1}{2},\frac{1}{2})}&=&\frac{1}{4}e^{A\kappa}(1)
\left[W_{{\rm vector}}(1)\right]^{AB}e^B_\kappa(0)
=\frac{1}{4}\left[W_{{\rm vector}}(1)\right]^{AA}\\
\la{gravWL4}
&=&\frac{1}{2}\Tr \,W^\pi\cdot
\frac{1}{2}\Tr \,W^\rho.
\eea
Its generalization to the holonomy in arbitrary representation
$(J_1,J_2)$ is obvious:

\beq
W^G_{(J_1,J_2)}=W^\pi_{J_1}\cdot W^\rho_{J_2},\qquad
W^{\pi,\rho}_J=\frac{1}{2J+1}\Tr_{(2J+1)}\,W^{\pi,\rho}.
\la{gravWL41}\eeq
Thus, in curved $d=4$ space the holonomy in the $(J_1,J_2)$ representation
is equal to a product of two Yang--Mills Wilson loops where the role of
Yang--Mills potentials is played by self-dual ($\pi^a_\mu$) and
anti-self-dual ($\rho^a_\mu$) parts of the spin connection. In section
9 we show that both $W^\pi$ and $W^\rho$ can be written in terms
of the metric tensor.

\section{Small Wilson loops}

For small-area contours the `gravitational Wilson loop' can be expanded
in powers of the area. The most straightforward way to do it is to use
the path-ordered form of $W^G$ as given by \eq{wlgsymb}. We take
a square contour of size $a\times a$ lying in the $12$ plane, and
expand the path-ordered exponent in powers of $a$. After some simple
algebra we obtain the first nontrivial term of that expansion which
happens to be $O(a^4)$:

\beq
W^G_{{\rm vector}}=\frac{1}{d}\left[W^G_{{\rm vector}}\right]^\kappa_\kappa
=1+\frac{a^4}{d}R^\kappa_{\;\lambda 12}R^\lambda_{\;\kappa 12}
=1-\frac{2(\Delta S)^{\mu\nu}(\Delta S)^{\mu^\prime \nu^\prime}}{4d}
R_{\kappa\lambda\mu\nu}R_{\rho\sigma\mu^\prime \nu^\prime}
g^{\kappa\rho}g^{\lambda\sigma},
\la{sWL1}\eeq
where $(\Delta S)^{\mu\nu}$ is the element of the surface lying in
the $(\mu\nu)$ plane. Notice that the first correction to the holonomy
is negative-definite. It should be stressed that the first-order term in
$\Delta S$ is, generally, present in the expansion of the parallel
transporter, however it vanishes when one takes the trace owing to the
identity $R^\kappa_{\;\kappa\mu\nu}\equiv 0$, so that the expansion for
the trace starts from the $(\Delta S)^2$ term.

In $d=3$ \eq{sWL1} can be further simplified because the Riemann tensor
is directly related to the Ricci tensor:

\beq
R_{ijkl}=R_{ik}g_{jl}-R_{il}g_{jk}+R_{jl}g_{ik}-R_{jk}g_{il}+
\frac{R}{2}(g_{il}g_{jk}-g_{ik}g_{jl}).
\la{RR}\eeq
Since Riemann tensor is antisymmetric inside both pairs of subscripts
we can replace

\beq
g^{km}g^{ln}\rightarrow \frac{1}{2}(g^{km}g^{ln}-g^{kn}g^{lm})
=\frac{1}{2g}\epsilon^{kli}\epsilon^{mnj}\,g_{ij}.
\la{r1}\eeq
We introduce the dual element of the surface,

\beq
\Delta S^{pq}=\epsilon^{pqr}\Delta S_r,
\la{dualS}\eeq
and have

\beq
\epsilon^{kli}\epsilon^{pqr}\,R_{klpq}=
-4\left(R^{ir}-\frac{1}{2}R\,g^{ir}\right),
\la{Einst}\eeq
which as a matter of fact is the Einstein tensor.
For the parallel transporter of arbitrary spin $J$
the factor ``$2$'' in the numerator of \ur{sWL1} should be replaced by
$J(J+1)$.

Combining all the factors we obtain

\beq
W^G_J=1-\frac{2J(J+1)}{3g}
\left(R^{ir}-\frac{1}{2}R\,g^{ir}\right)\;g_{ij}\;
\left(R^{js}-\frac{1}{2}R\,g^{js}\right)\Delta S_r\,\Delta S_s\,.
\la{sWL2}\eeq

This is our final expression for the trace of the parallel transporter
of spin $J$ in a curved $d=3$ space, for small loops. Notice that
\eq{sWL2} is invariant under diffeomorphisms.

\section{Gravitational Wilson loop in two dimensions}

In curved $d=2$ space the trace of the parallel transporter along
a closed loop can be computed exactly for any metric and presented
in the form of a `Stokes theorem'. The result is related to the
Gauss--Bonnet theorem and is generally known: we present it here
for the sake of completeness.

The key observation is that in two dimensions the spin connection
\ur{conn} has only one component,

\beq
\omega^{ab}_i=\epsilon^{ab}\,\omega_i.
\la{conn2}\eeq
In this section all indices assume only two values $1,2$.
According to \eq{partra2} the parallel transporter of a vector is
determined by the equation

\beq
\frac{dc^a}{d\tau}-\frac{dx^i}{d\tau}\omega_i\,\epsilon^{ab}\,c^b=0
\la{partra21}\eeq
whose solution is

\beq
c^a(\tau)=W^{ab}(\tau)\,c^b(0),\qquad
W^{ab}(\tau)=\left(\begin{array}{cc}
\cos \gamma(\tau)&\sin\gamma(\tau)\\
-\sin\gamma(\tau)&\cos\gamma(\tau)\end{array}\right),
\qquad\gamma(\tau)=\int_0^\tau d\tau\,\frac{dx^i}{d\tau}\omega_i.
\la{sol}\eeq
According to the general theorem of section 6 the gravitational Wilson
loop is equal to the Yang--Mills one, and we obtain

\beq
W^G_1=\frac{1}{2}W^{aa}(1)=\cos\Phi
\la{WL2d1}\eeq
where
\beq
\Phi=\gamma(1)=\int_0^1d\tau\frac{dx^i}{d\tau}\,\omega_i=
\frac{1}{2}\oint dx^i\,\epsilon_{ab}\,\omega^{ab}_i.
\la{gamm1}\eeq
This formula is not fully satisfactory yet since the holonomy is
expressed through the spin connection and not through the metric. It
can be achieved if we apply the Stokes theorem and write \eq{gamm1} in
a surface form. We have

\beq
\Phi=\frac{1}{2}\int dS\,\epsilon_{ab}\,\epsilon^{ij}\,
\partial_i\omega^{ab}_j
\la{gamm2}\eeq
where $dS$ is the element of the surface spanned on the contour.
Introducing the field strength related to the Riemann tensor,

\beq
F^{ab}_{ij}=\partial_i\omega^{ab}_j-\partial_j\omega^{ab}_j+
\omega^{ac}_i\omega^{cb}_j -\omega^{ac}_j\omega^{cb}_i
=R^{kl}_{\;\;ij}\,e^a_k\,e^b_l,\qquad\epsilon_{ab}\,e^a_k\,e^b_l
=\epsilon_{kl}\,\sqrt{g},
\la{fstr}\eeq
and noticing that in $d=2$ the commutator term is zero, we rewrite
\eq{gamm2} as

\beq
\Phi=\frac{1}{2}\int dS\,\sqrt{g}\,R,\qquad\qquad W^G_1=\cos\Phi,
\la{gamm3}\eeq
where $R=(1/2)\,\epsilon^{ij}\,\epsilon_{kl}\,R^{kl}_{ij}$ is the
scalar curvature. It is gratifying that the holonomy is expressed
through the Einstein--Hilbert action, known to be a full derivative in
two dimensions. Needless to explain, \eq{gamm3} is
diffeomorphism-invariant.

In $d=2$ there is essentially only one component of the Riemann tensor,

\beq
R_{1212}=\frac{1}{2}\,R\,g
\la{Riem2}\eeq
(see \cite{LogL}). Taking it into account it is easy to check
that for small areas the expansion of \eq{gamm3} gives the same
result as \eq{sWL1} written for small loops.

\section{An example of big loops: constant curvature background in $d=3$}

In three dimensions the Riemann tensor is expressible through the
Ricci tensor, see \eq{RR}. Because of it, diffeomorphism-invariant
information about curved spaces is fully contained in the three
eigenvalues of the symmetric Ricci tensor,

\beq
R^i_j=\lambda\,\delta^i_j,
\la{eig1}\eeq
the scalar curvature being the sum of the three,
$R=\lambda_1+\lambda_2+\lambda_3$. For example, the de~Sitter $S^3$
space corresponds to $\lambda_1=\lambda_2=\lambda_3=R/3={\rm const.}$
In this section we would like to consider another special case
of constant curvature, namely the {\em cylinder space} $S^2\times R$,
characterized by $\lambda_1=\lambda_2=R/2={\rm const},\quad
\lambda_3=0$. We shall see that the parallel transporter in
such spaces can be computed for any form of the contour and any metric
and that the gravitational Wilson loop is given by an elegant formula.

A general metric can be considered as induced by 6 external
coordinates $w^A(x_1,x_2,x_3)$:

\beq
g_{ij}=\partial_iw^A\partial_jw^A,\qquad A=1,\ldots,6.
\la{genmetr}\eeq
In the special case of the cylinder space $S^2\times R$ it is
sufficient to use only four external coordinates
$w^a\quad(a=1,2,3)$ and $w^4$, subject to the constraint

\beq
\sum_{a=1}^3(w^a)^2=\frac{2}{R}.
\la{constr1}\eeq
An example of such external coordinates is given by

\beq
w^{1,2,3}(x)=\sqrt{\frac{2}{R}}\,\frac{x^{1,2,3}}{r},\qquad
w^4(x)=\sqrt{\frac{2}{R}}\;\ln r,
\la{ex1}\eeq
leading to the metric tensor

\beq
g_{ij}=\frac{2}{R}\;\frac{1}{r^2}\;\delta_{ij},\qquad\sqrt{g}=
\left(\frac{2}{R}\right)^{\frac{3}{2}}\;\frac{1}{r^3}.
\la{exm1}\eeq
A simple calculation using formulae from section 3 shows that this
metric indeed gives one eigenvalue of the Ricci tensor zero and the
other two equal to a constant $R/2$. Since the eigenvalues of the Ricci
tensor are diffeomorphism-invariant, a general change of coordinates,
$x^i\to y^i(x)$, in \eq{ex1} result in the same eigenvalues.
Therefore, the most general description of the cylinder spaces
$S^2\times R$ is given by
\bea
\la{wgen}
w^a(x)&=&\sqrt{\frac{2}{R}}\frac{y^a(x)}{|y(x)|},\qquad
w^4(x)=\sqrt{\frac{2}{R}}\;\ln |y(x)|,\qquad
g_{ij}=\frac{2}{R}\frac{\partial_i y^a\partial_i y^a}{y^2},\\
\sqrt{g}&=&\left(\frac{2}{R}\right)^{\frac{3}{2}}\,
\frac{1}{3!}\,\epsilon^{ijk}\,\epsilon_{abc}\,
\frac{\partial_iy^a\partial_jy^b\partial_ky^c}{|y|^3}
=\frac{1}{2}\sqrt{\frac{R}{2}}\,\epsilon^{ijk}\,\epsilon_{abc}\,
\partial_iw^a\,\partial_jw^b\,w^c\,\partial_kw^4,
\la{ggen}\eea
where $y^a(x)$ are three arbitrary functions of coordinates $x^i$.
Notice that $g_{ij}$ is given by a product of two matrices
$M^a_i=\partial_iy^a/|y|$ and hence $\sqrt{g}$ is itself a determinant
(of the matrix $M$).

Our aim is to calculate the Wilson loop for any contour in any metric
\ur{wgen} corresponding to the cylinder spaces. We shall make use of
the fact that the Wilson loop is diffeomorphism-invariant. If we
compute it for a general contour in some metric representing
cylinder spaces, the most general case is recovered by diffeomorphisms
of both the contour and the metric. We shall start with the specific
metric given by \eqs{ex1}{exm1}.

Given the metric tensor \ur{exm1}, we construct a vielbein
corresponding to it. This is, of course, not unique but any choice
of the vielbein will suit us. We choose

\beq
e^a_i=\sqrt{\frac{2}{R}}\;\frac{1}{r}\;\delta^a_i,\qquad
e^a_ie^a_j=g_{ij}.
\la{exv1}\eeq
Given the vielbein we construct the spin connection from its
definition \ur{conn} or the Yang--Mills field and obtain

\beq
A^a_i=-\frac{1}{2}\epsilon^{abc}\omega^{bc}_i
=\epsilon^{aij}\;\frac{x^j}{r^2},
\la{mon1}\eeq
which happens to be the field of the Wu--Yang monopole;
the scalar curvature $R$ has dropped from the spin connection.
According to the theorem of section 4 the gravitational Wilson
loop is equal to the Yang--Mills Wilson loop, provided the Yang--Mills
potential $A^a_i$ is the spin connection of the metric under
consideration. Therefore, all we have to do is to compute the Wilson loop
in the field of the Wu--Yang monopole, but for a general contour.

This task is easily solvable if we make use of another invariance, this
time it is gauge invariance of the Wilson loop. It is well known that
the Wu--Yang monopole in the hedgehog gauge \ur{mon1} can be
transformed to the string gauge where the potential has only
one nonzero component along the third colour axis (plus a Dirac
string). In this gauge the Yang--Mills potential is basically abelian
so that one has for the Wilson loop in any representation $J$:

\beq
W^G_J=W^{YM}_J=\frac{1}{2J+1}\sum_{m=-J}^J\exp\, im\Phi,\qquad
\Phi=\oint dx^i A^3_i=\int dS_i\; \frac{x^i}{r^3}.
\la{WLmon1}\eeq
In the last equation we have used the normal Stokes theorem for the
circulation and also the fact that in the string gauge the magnetic
field of the monopole is the Coulomb field of a point charge; $dS_i$ is
the element of the surface spanned on the contour, and is
orthogonal to the surface.

\Eq{WLmon1} is the gravitational Wilson loop for arbitrary contours but
in a specific metric given by \eq{exm1}. In order to generalize it to
a general metric \ur{wgen} all one needs is to perform the general
coordinate transformation of \eq{WLmon1}. To this end it is convenient
to use, instead of $dS_i$, its dual $dS^{ij}$ such that
$dS_i=\epsilon_{ijk}\;dS^{jk}$. Taking into account that under the
general coordinate transformation $x^i\to y^i(x)$ the contravariant
vector transforms as $V^i\to V^k\partial_ky^i$, and the antisymmetric
contravariant tensor transforms as
$dS^{ij}\to dS^{mn}\,\partial_my^i\,\partial_ny^j$, we get for the flux
in \eq{WLmon1}:

\beq
\Phi=\int dS_i\; \frac{x^i}{r^3} =
\int dS^{ij}\;\epsilon_{ijk}\; \frac{x^k}{r^3} \rightarrow
\int dS^{mn}\;\frac{\epsilon_{ijk}\;\partial_my^i\,\partial_ny^j\,y^k}
{|y|^3}.
\la{fluxmon1}\eeq
This equation takes a more symmetric form if one uses the external
coordinates \ur{wgen}:

\beq
\Phi=\left(\frac{2}{R}\right)^{\frac{3}{2}}\,\frac{1}{2}
\int dS_k\;\epsilon_{abc}\;\epsilon^{ijk}\;
\partial_iw^a\,\partial_jw^b\,w^c,
\qquad\sum_{a=1}^3w^{a2}=\frac{2}{R}.
\la{fluxmon2}\eeq
\Eq{WLmon1} together with \eq{fluxmon2} is our final result for the
gravitational Wilson loop in the cylinder $S^2\times R$ space of
constant curvature R. Through \eq{wgen} the
Wilson loop implicitly depends on the metric. Let us make a few
comments.
\begin{itemize}

\item The parallel transporter should depend on the metric along the
contour but not on the surface spanned on the contour as it can
be drawn arbitrarily. Despite the surface form of the result it is
indeed so due to the fact that

\beq
\partial_k\left(\epsilon_{abc}\;\epsilon^{ijk}\;
\partial_iw^a\,\partial_jw^b\,w^c\right)=0.
\la{div1}\eeq
Therefore, the flux in \eq{fluxmon2} can be presented as a
circulation of a certain vector.

\item The flux \ur{fluxmon2} coincides in form with a well-known
expression for the winding number of the mapping $S^2\mapsto S^2$. For
a closed or infinite surface the winding number is normalized as

\beq
\frac{1}{8\pi}\left(\frac{2}{R}\right)^{\frac{3}{2}}
\int dS_k\;\epsilon_{abc}\;\epsilon^{ijk}\;
\partial_iw^a\,\partial_jw^b\,w^c=Q=\;{\rm integer}.
\la{wind1}\eeq

\item For small contours \eqs{WLmon1}{fluxmon2} reproduce the result of
the previous section. To check it, let us rewrite the general
small-loop expansion \ur{sWL1} for the concrete metric \ur{ex1}. We
find:

\beq
R_{klpq}=\frac{2}{R\,r^6}\;\epsilon_{klu}\,x^u\,\epsilon_{pqv}x^v,
\qquad g^{ij}=\frac{R}{2}\,r^2\,\delta^{ij}.
\la{Ricci3}\eeq
Putting it into \eq{sWL1} and then performing a general coordinate
transformation $x^i\to y^i(x)$ we obtain after simple algebra

\beq
W^G_J=1-\frac{J(J+1)}{6}\left(\frac{\epsilon_{klu}\,y^u\,
\partial_iy^k\,\partial_jy^l\;\Delta S^{ij}}{|y|^3}\right)^2
\la{smWL3}\eeq
which coincides exactly with the expansion of \eq{WLmon1} in small
loop areas $\Delta S$ up to the second order.

\end{itemize}

\section{Non-Abelian Stokes theorem in $d=3$ gravity}

In section 4.1 we have shown that the gravitational Wilson loop in
integer representations $J$ as a functional of the metric is equal to
the Yang--Mills Wilson loop as a functional of the Yang--Mills
potential, provided one takes this potential equal to the spin
connection corresponding to the metric in question, see \eq{YMc}.
For holonomies with half-integer $J$ only the representation via spin
connection is available. In this section we present the holonomy
both for integer and half-integer $J$ in a unified way, by introducing
the `non-Abelian Stokes theorem' similar to that for the Yang--Mills
Wilson loop. We shall show that, being written down in a surface
form, holonomies for both integer and half-integer $J$ are actually
expressible through the metric tensor and its derivatives only: the
vielbein and spin connection (not uniquely defined by the metric) are,
in fact, unnecessary.

For any representation $J$ we can use our non-Abelian Stokes
formula \ur{W3} where the Yang--Mills potential is directly related
to the spin connection:
\bea
W^G_J&=&W^{YM}_J[{\rm spin\;connection}]\\
\nonumber
&=&\int D{\bf n}\,\delta({\bf n}^2-1)\,\exp\frac{iJ}{2}\int\,
d^2 S^{ij}\left[-F_{ij}^an^a
+\epsilon^{abc} n^a\left(D_i n\right)^b \left(D_j n\right)^c\right]\,.
\la{W22}\eea
We shall replace the element of the surface by its dual, $dS^{ij}=
\epsilon^{ijp}\,dS_p$. Our goal will be to rewrite this representation
of the Wilson loop in terms of the metric of the curved $3d$ space. To
this end we first decompose the integration unit vector $\n$ in the
dreibein:

\beq
n^a=m^i\,e^a_i,\qquad n^an^a=m^im^je^a_ie^a_j=m^im^jg_{ij}=1.
\la{dec1}\eeq
The new 3-vector ${\bf m}$ is a {\em covariantly} unit vector. Since the
background metric $g_{ij}$ is fixed we just change the integration
variables from ${\bf n}$ to ${\bf m}$, the new integration measure being

\beq
\int D{\bf n}\,\delta({\bf n}^2-1)\ldots
=\int D{\bf m}\,\sqrt{g}\,\delta(m^im^j\,g_{ij}-1)\ldots
\la{covmeas}\eeq

We next use the relation \ur{rel5} of the field strength $F^a_{ij}$
computed from the spin connection $A^a_i=(1/2)\epsilon^{abc}\omega^{bc}_i$,
to the Riemann tensor. The first term in the exponent of \eq{W22} becomes

\beq
{\rm first\;term}= -dS_p\epsilon^{ijp}\,\left(-\frac{1}{2}\right)\,
\epsilon^{abc}\,m^ne^a_n\,R^l_{\;kij}\,e^b_l\,e^{ck}.
\la{ft1}\eeq
Using

\beq
\epsilon^{abc}\,e^{bl}\,e^{ck}=\frac{1}{\sqrt{g}}\epsilon^{lkm}\,e^a_m,
\qquad \sqrt{g}=\det\,e^a_i,
\la{rel4}\eeq
\eq{ft1} can be continued as

\beq
{\rm first\;term}= dS_p\epsilon^{ijp}\,\frac{1}{2\sqrt{g}}
\,R_{ijkl}\,\epsilon^{klm}\,g_{mn}m^n.
\la{ft2}\eeq
The combination of the covariant Riemann tensor and two antisymmetric
epsilons has been encountered in section 5: in $d=3$ it gives the
Einstein tensor, see \eq{Einst}. We get therefore

\beq
{\rm first\;term}=
dS_p\,\sqrt{g}\,\left(R\delta^p_n-2R^p_n\right)\,m^n,
\la{ft3}\eeq
where $R^p_n$ is the Ricci tensor and $R=R^k_k$ is the scalar
curvature.

We now turn to the second term in the exponent of \eq{W22} and again
use the decomposition \ur{dec1}. We exploit the fundamental relation
\ur{e1} which can be presented as

\beq
D^{bb^\prime}_jn^{b^\prime}=e^b_k\left(\nabla_{\!j}\right)^k_lm^l
\la{e3}\eeq
where $D^{bb^\prime}_j=\partial_j\delta^{bb^\prime}+
\epsilon^{bcb^\prime}A^c_j$ is the Yang--Mills and
$(\nabla_{\!j})^k_l=\partial_j\delta^k_l+\Gamma^k_{jl}$ is the
gravitational covariant derivatives. Therefore, the second term is
\bea
{\rm second\;term}&=&dS_p\,\epsilon^{abc}e^a_ke^b_le^c_n\,
\epsilon^{ijp}\,m^k(\nabla_{\!i})^l_{l^\prime}m^{l^\prime}
(\nabla_{\!j})^n_{n^\prime}m^{n^\prime}
\nonumber\\
\nonumber\\
&=&dS_p\,\sqrt{g}\,\epsilon^{ijp}\,\epsilon_{kln}\,
m^k(\nabla_{\!i}\,m)^l(\nabla_{\!j}\,m)^n.
\la{st1}\eea

Gathering \eqss{covmeas}{ft3}{st1} together we get finally a
non-Abelian Stokes theorem for the gravitational Wilson loop
or the trace of the parallel transporter for spin $J$ along a closed
contour:
\bea
W^G_J&=&\int D{\bf m}\,\sqrt{g}\;\delta(m^im^j\,g_{ij}-1)
\nonumber\\
\nonumber\\
&\times&\exp\;i\frac{J}{2}\int\! dS_k\,\sqrt{g}\,
\left[\left(R\delta^k_p-2R^k_p\right)\,m^p+\epsilon^{ijk}\,
\epsilon_{pqr}\,m^p(\nabla_{\!i}\,m)^q(\nabla_{\!j}\,m)^r\right].
\la{gravWL1}\eea

Several comments are in order here.
\begin{itemize}
\item The holonomy being defined as a path-ordered exponent is
expressed here by a simple exponent of a integral over the surface
spanned on the closed contour. That is why we call it a `Stokes
theorem'. The price to pay is a functional integration over covariantly
unit vector ${\bf m}$ defined on the surface.

\item \Eq{gravWL1} is {\em invariant under diffeomorphisms}, in the
sense that if one makes a general coordinate transformation $x^i\to
x^{\prime\,i}(x^i)$ and changes the surface appropriately, the holonomy
remains invariant.

\item The parallel transporter depends only on the contour
but should not depend on the way one spans a surface on that contour.
The surface integral in \eq{gravWL1} has the form

\beq
\int dS_k \;\sqrt{g} \;V^k,
\la{surf1}\eeq
and the condition that it does not depend on the form of the surface is

\beq
\partial_k\left(\sqrt{g}\,V^k\right)=0,
\la{div3}\eeq
being equivalent to the condition

\beq
\left(\nabla_{\!k}\right)^k_lV^l=0,
\la{div2}\eeq
since $\Gamma^k_{kl}=\Gamma^k_{lk}=\partial_l\ln\sqrt{g}$. The check
of \eq{div2} is rather lengthy and we relegate it to the Appendix.

\item Condition \ur{div2} or equivalently \ur{div3} being satisfied
means that the surface integral can be written as

\beq
\int dS_k\;\sqrt{g}\,V^k=\int dS_k\epsilon^{ijk}\partial_jB_k
=-\oint dx^i\,B_i
\la{rotor}\eeq
proving that it depends only on the contour, as it should be.
However, the vector field $B_i$ cannot be uniquely determined from
the metric tensor and the covariantly unit vector ${\bf m}$.

\item The following comment is closely related to the previous.
Parallel transporters of integer spins 1,2,... are defined via
Christoffel's $\Gamma$ symbols and hence by the metric tensor,
while parallel transporters of half-integer spins $1/2,\;3/2,...$
are not: they are defined via the spin connection which is not uniquely
constructed from the metric. Nevertheless, it should be expected that
the holonomy for half-integer spins, i.e. the trace of a parallel
transporter along a closed loop, being a diffeomorphism-invariant
quantity, should be expressible through the metric only. The above
\eq{gravWL1} solves this non-trivial problem: only the metric and
its derivatives are involved.

The solution is possible only when one presents the holonomy in the
form of a surface integral, as in \eq{gravWL1}. One cannot do it in a
contour form as it is not uniquely expressible through the metric. Were
that possible, one would be able to write down a parallel transporter
in terms of metric along an open contour as well, but that is not so
for half-integer spins.

\item \Eq{gravWL1} solves another long-standing problem, this time
in the Yang--Mills theory. In another paper \cite{DP4} we have shown
that the $SU(2)$ Yang--Mills partition function in three dimensions can
be exactly rewritten in terms of gauge-invariant quantities which
happen to be the six components of the metric tensor of the dual
space.  The usual argument why such rewriting is not too useful is that
external sources couple to the Yang--Mills potential and not to
gauge-invariant quantities. However, now we have demonstrated that a
typical source, i.e. the Yang-Mills Wilson loop can be expressed not
only through the potential but also through the metric tensor which is
gauge-invariant. Thus, not only the partition function but
also Wilson loops in the $d=3$ Yang--Mills theory can be expressed
through local gauge-invariant quantities.
\end{itemize}

\section{Non-Abelian Stokes theorem in $d=4$ gravity}

The aim of this section is to present the holonomy $W^G_{(J_1,J_2)}$
in curved $d=4$ space in the representation $(J_1,J_2)$ in terms of the
metric tensor and its derivatives. \Eq{gravWL41} presents the holonomy
in terms of the (anti)self-dual parts of the spin connection not
uniquely determined by the metric, which is not satisfactory. In
addition, we would like to get rid of the path-ordering in the
Yang--Mills Wilson loops $W^{\pi,\rho}$ entering \eq{gravWL41}. Both
goals are achieved via the non-Abelian Stokes theorem similar to that
of the previous section, which we are going to derive now.

We start by applying the representation \ur{W3} to the Yang--Mills
Wilson loop $W^\pi$:

\beq
W^\pi_{J}=\int D{\bf n}\,\delta({\bf n}^2-1)\,\exp\,i\frac{J}{2}
\int dS^{\mu\nu}\left[-F^a_{\mu\nu}(\pi)n^a+
\epsilon^{abc}n^a\left(D_\mu(\pi)n\right)^b
\left(D_\nu(\pi)n\right)^c\right]
\la{Wpi1}\eeq
where
$D^{ab}_\mu(\pi)=\partial_\mu\,\delta^{ab}+\epsilon^{acb}\,\pi^c_\mu$
is the covariant derivative with respect to the self-dual part of the
spin connection and $F^a_{\mu\nu}(\pi)$ is the appropriate field
strength \ur{fspi}; it is related to the Riemann tensor via \eq{FpitoR}.
Let us introduce an antisymmetric tensor

\beq
m^{\kappa\lambda}=\frac{1}{2}n^a\,\eta^{aAB}\,e^{A\kappa}e^{B\lambda}.
\la{m1}\eeq
The first term in \eq{Wpi1} can be written as
$-R_{\kappa\lambda\mu\nu}m^{\kappa\lambda}$. The tensor
$m^{\kappa\lambda}$ has actually only two independent components. To
see this we introduce two covariant projector operators,
\bea
\la{Pplus}
P^+_{\kappa\lambda\mu\nu}&=&\frac{1}{4}\eta^{aAB}\eta^{aCD}
e^A_\kappa e^B_\lambda e^C_\mu e^D_\nu=
\frac{1}{4}(g_{\kappa\mu}g_{\lambda\nu}
-g_{\kappa\nu}g_{\lambda\mu}+\sqrt{g}\epsilon_{\kappa\lambda\mu\nu}),\\
\nonumber\\
\la{Pminus}
P^-_{\kappa\lambda\mu\nu}&=&\frac{1}{4}\bar\eta^{aAB}\bar\eta^{aCD}
e^A_\kappa e^B_\lambda e^C_\mu e^D_\nu=
\frac{1}{4}(g_{\kappa\mu}g_{\lambda\nu}
-g_{\kappa\nu}g_{\lambda\mu}-\sqrt{g}\epsilon_{\kappa\lambda\mu\nu}),
\eea
satisfying projector conditions,
\bea
\la{P1}
P^\pm_{\kappa\lambda\mu\nu}\, g^{\mu\mu^\prime}g^{\nu\nu^\prime}
P^\pm_{\mu^\prime\nu^\prime\rho\sigma}
&=&P^\pm_{\kappa\lambda\rho\sigma},\\
\la{P2}
P^\pm_{\kappa\lambda\mu\nu}\, g^{\mu\mu^\prime}g^{\nu\nu^\prime}
P^\mp_{\mu^\prime\nu^\prime\rho\sigma}
&=&0,\\
\la{P3}
P^\pm_{\kappa\lambda\mu\nu}\, g^{\kappa\mu}g^{\lambda\nu}
&=&3.
\eea
$P^\pm_{\kappa\lambda\mu\nu}$ are (covariantly) orthogonal
projectors, each having 3 zero and 3 nonzero eigenvalues. They project
a general antisymmetric tensor into (covariantly) self-dual and
anti-self-dual parts, respectively. It is easy to check that the tensor
$m^{\kappa\lambda}$ introduced in \eq{m1} is self-dual,

\beq
P^-_{\kappa\lambda\mu\nu}\,m^{\kappa\lambda}=0,
\la{m11}\eeq
and satisfies the normalization,

\beq
m^{\kappa\lambda}\,m_{\kappa\lambda}=
P^+_{\kappa\lambda\mu\nu}\,m^{\kappa\lambda}\,m^{\mu\nu}=1,
\la{m12}\eeq
being a consequence of the normalization ${\bf n}^2=1$. Therefore,
$m^{\kappa\lambda}$ has, indeed, only two independent degrees of
freedom in a given metric. We change the integration variables
in \eq{Wpi1} from ${\bf n}$ to $m^{\kappa\lambda}$:

\beq
\int D{\bf n}\,\delta({\bf n}^2-1)\ldots=
\int Dm^{\kappa\lambda}\,\sqrt{g}\,\delta(
P^-_{\kappa\lambda\mu\nu}\,m^{\mu\nu})\,
\delta(m^{\kappa\lambda}m_{\kappa\lambda}-1)\ldots
\la{mm1}\eeq

Let us now compute the covariant derivative of $m^{\kappa\lambda}$:
\bea
\nonumber
m^{\kappa\lambda}_{\;\;\;;\mu}&=&
\partial_\mu m^{\kappa\lambda}+\Gamma^\kappa_{\mu\nu}m^{\nu\lambda}
+\Gamma^\lambda_{\mu\nu}m^{\kappa\nu}\\
\nonumber
&=&\frac{1}{2}\eta^{aAB}\left[\partial_\mu n^a\,e^{A\kappa}e^{B\lambda}
+n^a(\partial_\mu
e^{A\kappa}+\Gamma^\kappa_{\mu\nu}e^{A\nu})e^{B\lambda}
+n^a e^{A\kappa}(\partial_\mu
e^{B\lambda}+\Gamma^\lambda_{\mu\nu}e^{B\nu})\right]\\
\la{m;1}
&=&\frac{1}{2}\eta^{aAB}\left[\partial_\mu n^a\,e^{A\kappa}e^{B\lambda}
-n^a\omega^{AC}_\mu e^{C\kappa}e^{B\lambda}
-n^a e^{A\kappa}\omega^{BC}_\mu e^{C\lambda}\right]
\eea
where in the last equation we have used the fundamental relation
\ur{e1}. We now insert the decomposition of the spin connection
$\omega^{AB}_\mu$ into self-dual and anti-self-dual parts, \eq{omdec}.
Using the relations for the $\eta,\bar\eta$ symbols,
\bea
\la{eta2}
\eta^{aAB}\eta^{bAC}&=&\delta^{ab}\delta^{BC}+\epsilon^{abc}\eta^{cBC},
\qquad
\bar\eta^{aAB}\bar\eta^{bAC}=\delta^{ab}\delta^{BC}+\epsilon^{abc}
\bar\eta^{cBC},\\
\la{eta3}
\eta^{aAB}\bar\eta^{bAC}&=&\eta^{aAC}\bar\eta^{bAB},
\eea
it is easy to see that only the self-dual piece of $\omega^{AB}_\mu$
survives in \eq{m;1}, giving

\beq
m^{\kappa\lambda}_{\;\;\;;\mu}=
\frac{1}{2}\eta^{aAB}e^{A\kappa}e^{B\lambda}\left(\partial_\mu\,\delta^{ab}
+\epsilon^{acb}\pi^c_\mu\right)\,n^b=
\frac{1}{2}\eta^{aAB}e^{A\kappa}e^{B\lambda}\left(D_\mu(\pi)\,n\right)^a.
\la{m;2}\eeq
In other words, the gravitational covariant derivative of
$m^{\kappa\lambda}$ is expressed through the Yang--Mills covariant
derivative of the ${\bf n}$ field, which is encountered in the second
term of \eq{Wpi1}.

Using consecutively \eqs{eta2}{m;2} we obtain after simple algebra
the final expression for $W^\pi_J$ \ur{Wpi1} but this time presented
in terms of the metric:
\bea
\nonumber
W^\pi_{J_1}&=&
\int Dm^{\kappa\lambda}\,\sqrt{g}\;\delta(
P^-_{\kappa\lambda\mu\nu}\,m^{\mu\nu})\,
\delta(m^{\kappa\lambda}m_{\kappa\lambda}-1)\\
\la{Wpi2}
&\times& \exp\,i\frac{J_1}{2}\int
dS^{\mu\nu}\left[-R_{\kappa\lambda\mu\nu}\,
m^{\kappa\lambda}-\frac{1}{2}\,\sqrt{g}\,\epsilon_{\kappa\rho\sigma\tau}\,
g_{\lambda\lambda^\prime}\,m^{\kappa\lambda^\prime}\,
m^{\lambda\rho}_{\;\;\; ;\mu}\,m^{\sigma\tau}_{\;\;\; ;\nu}\right].
\eea
Similarly, $W^\rho$ is obtained by integrating over anti-self-dual
covariantly unit tensors:
\bea
\nonumber
W^\rho_{J_2}&=&
\int Dm^{\kappa\lambda}\,\sqrt{g}\;\delta(
P^+_{\kappa\lambda\mu\nu}\,m^{\mu\nu})\,
\delta(m^{\kappa\lambda}m_{\kappa\lambda}-1)\\
\la{Wrho2}
&\times& \exp\,i\frac{J_2}{2}\int
dS^{\mu\nu}\left[-R_{\kappa\lambda\mu\nu}\,
m^{\kappa\lambda}+\frac{1}{2}\,\sqrt{g}\,\epsilon_{\kappa\rho\sigma\tau}\,
g_{\lambda\lambda^\prime}\,m^{\kappa\lambda^\prime}\,
m^{\lambda\rho}_{\;\;\; ;\mu}\,m^{\sigma\tau}_{\;\;\; ;\nu}\right].
\eea
As derived in section 4.2, the gravitational holonomy in representation
$(J_1,J_2)$ is the product of the two,

\beq
W^G_{(J_1,J_2)}=W^\pi_{J_1}W^\rho_{J_2}.
\la{WG4}\eeq
\Eqs{Wpi2}{Wrho2} and \ur{WG4} form the `non-Abelian Stokes theorem'
for the holonomy in curved $d=4$ space. It expresses the holonomy
via surface integrals spanned on the contour, and presents it
in terms of the metric tensor and its derivatives only, without
reference to the spin connection, even for half-integer representations
$J_1,J_2$.

\section{Conclusions}

The main result of this paper are the non-Abelian Stokes theorems
for the holonomies: the Yang--Mills Wilson loop (\eq{W3}) and the
traces of parallel transporters in curved $d=3$ (\eq{gravWL1})
and $d=4$ (\eqs{Wpi2}{Wrho2}) spaces.
In all cases the path-ordered exponents of the connections are removed
and replaced by ordinary exponents of surface integrals which, however,
do not actually depend on the way the surface is spanned on the
contour. The price to pay for the removal of path ordering is high:
we obtain functional integrals instead. In the simplest case of the
$SU(2)$ Yang--Mills theory it is an integral over a unit 3-vector
${\bf n}$ `living' on the surface, in the case of the $d=3$
Riemannian manifold it is an integral over a covariantly unit
3-vector ${\bf m}$, in case of $d=4$ one integrates over
(anti)self-dual covariantly unit tensors.

In spite of functional integration we believe that our formulae are
aesthetically appealing. As compared to path-ordered exponents they
are better suited to averaging over quantum ensembles of Yang--Mills
fields or over various metrics. We hope that elegant formulae can be
also used in more general settings.

In addition to the general non-Abelian Stokes formulae we have
presented holonomy as a surface integral for a specific background,
namely for a $d=3$ space of constant curvature with cylinder topology
$S^2\times R$. The `gravitational Wilson loop' is given by a formula
for the character whose argument is the winding number of external
coordinates, see section 8.

Parallel transporters of integer spins have a dual description: one
can define them either as a path-ordered exponent of Christoffel symbols
or as a path-ordered exponent of spin connection in appropriate
representation. In section 4 we have shown that both representations
are equivalent. Even though spin connection is not uniquely determined
by the metric tensor, this equivalence means that the holonomy written
in terms of spin connection can be in fact expressed through the
metric only.

For half-integer spins the situation is far less trivial since the
only way to define the holonomy is via the spin connection, and it is
not at all clear beforehand that it can be uniquely written through the
metric tensor and its derivatives. The non-Abelian Stokes theorem of
this paper demonstrates that it is indeed so, but only when one
presents the holonomy in the surface form which is uniquely determined
by the metric. Though the surface integral does not depend on the way
one draws the surface and can be actually written as an integral
along the contour, the contour form is not uniquely defined by the
surface one, and it reflects the ambiguity in determining the spin
connection from the metric.

This finding has interesting implication for Yang--Mills theory in three
dimensions, which can be identically reformulated as a quantum gravity
theory, with the partition function written as a functional integral
over the metric tensor of the dual space \cite{DP4}. This metric tensor
is local and gauge invariant (in the Yang--Mills sense). However, one
might wish to calculate the average of the Wilson loop which,
originally, is defined by the Yang--Mills potential, but not by the
metric tensor. In the `quantum gravity' formulation the Yang--Mills
Wilson loop becomes a parallel transporter in the
gravitational sense. Therefore, it is very important that it, too, is
expressible through the gauge-invariant metric tensor, in any
representation. Thus, not only the partition function but also the
Wilson loop can be presented in terms of {\em local} and {\em
gauge-invariant} quantities. This subject will be presented in more
details elsewhere.\\

One of us (V.P.) thanks NORDITA for kind hospitality and the
Russian Foundation for Basic Research for partial support, grant
RFBR-00-15-96610.

\section*{Appendix. Proof that \eq{gravWL1} does not depend on the
surface}

The path-integral representation for the `gravitational Wilson loop'
\ur{gravWL1} should not depend on the choice of the surface but only on
the contour on which the surface is spanned. To prove that it is indeed
so, one has to check \eq{div2},

\beq
\left(\nabla_{\!k}\right)^k_lV^l=0,
\la{div21}\eeq
where

\beq
V^k=\left(R\delta^k_p-2R^k_p\right)\,m^p+
\epsilon^{ijk}\,\epsilon_{pqr}\,m^p(\nabla_{\!i}\,m)^q(\nabla_{\!j}\,m)^r,
\qquad m^im^jg_{ij}=1.
\la{Vk}\eeq

To simplify notations we denote covariant derivatives
by ``;'' (see \cite{LogL}). Explicitly, the covariant derivatives of a
scalar, vector, tensor are given by
\bea
\nonumber
S_{;k}&=&\partial_k\,S,\\
\nonumber
V^i_{;k}&=&\partial_k\,V^i+\Gamma^i_{kl}\,V^l,\qquad
V_{i;k}=\partial_k\,V_i-\Gamma^l_{ik}\,V_l, \\
\la{covdiff}
T^{ij}_{;k}&=&\partial_k\,T^{ij}+\Gamma^i_{kl}\,T^{lj}
+\Gamma^j_{kl}\,T^{il},\qquad
T_{ij;k}=\partial_k\,T_{ij}-\Gamma^l_{ik}\,T_{lj}
-\Gamma^l_{jk}\,T_{il},\quad{\rm etc.}
\eea
An ordinary derivative of a convolution of two tensors can be written
as a sum of covariant derivatives:

\beq
\partial_k\,\left(T^{(1)\ldots i}_{\ldots}\,T^{(2)\ldots}_{\ldots i}
\right)=T^{(1)\ldots i}_{\ldots ;k}\,T^{(2)\ldots}_{\ldots i}
+T^{(1)\ldots i}_{\ldots}\,T^{(2)\ldots}_{\ldots i;k}\,.
\la{conv}\eeq

We apply the covariant derivative to the first term of the vector
$V^k$:

\beq
\nabla_{\!k}\left[\left(R\delta^k_p-2R^k_p\right)\,m^p\right]
=\left(R\delta^k_p-2R^k_p\right)_{;k}\,m^p
+\left(R\delta^k_p-2R^k_p\right)\,m^p_{;k}.
\la{ftd1}\eeq
The covariant derivative of the Einstein tensor is known to be zero
(\cite{LogL}, eq.(92.10)). Therefore, only the second term in \eq{ftd1}
survives.

Next, we apply the covariant derivative to the second term of the
vector $V^k$:

\[
\nabla_{\!k}\left[\epsilon^{ijk}\,\epsilon_{pqr}\,m^p(\nabla_{\!i}\,m)^q
(\nabla_{\!j}\,m)^r\right]
\]
\beq
=\epsilon^{ijk}\,\epsilon_{pqr}\,(\nabla_{\!k}\,m)^p(\nabla_{\!i}\,m)^q
(\nabla_{\!j}\,m)^r
+2\,\epsilon^{ijk}\,\epsilon_{pqr}\,m^p(\nabla_{\!i}\,m)^q
(\nabla_{\!k}\nabla_{\!j}\,m)^r.
\la{std1}\eeq
The first term here is zero, for the following reasons. We
differentiate the condition that $m^i$ is a covariantly unit vector,

\beq
0=\partial_k\left(m^im^jg_{ij}\right)=2\,g_{ij}
\left(\nabla_{\!k}\,m\right)^i\,m^j
=2\,\left(\nabla_{\!k}\,m\right)^i\,m_i,
\la{covun1}\eeq
since the covariant derivative of the metric tensor is zero. It means
that three vectors, $\left(\nabla_{\!1,2,3}\,m\right)^i$ are not
linearly independent as there cannot be three linearly independent
vectors orthogonal to some vector (in this case $m_i$) in three
dimensions.  The first term in \eq{std1} is an antisymmetrized product
of these three linearly-dependent vectors, therefore, it is zero.

As to the second term in \eq{std1} we notice that it contains the
commutator of covariant derivatives, equal to

\beq
\epsilon^{ijk}(\nabla_{\!k}\nabla_{\!j}\,m)^r
=\frac{1}{2}\epsilon^{ijk}\left[\nabla_{\!k}\nabla_{\!j}\right]^r_s\,m^s
=\frac{1}{2}\epsilon^{ijk}\,g^{rt}\,R_{tskj}\,m^s
\la{std2}\eeq
where $R_{tskj}$ is the Riemann tensor. Therefore, the second (and the
only nonzero) term in \eq{std1} can be written as

\beq
\epsilon^{ijk}\,\epsilon_{pqr}\,g^{rt}\,R_{tskj}\,
m^pm^s\left(\nabla_{\!i}\,m\right)^q.
\la{std3}\eeq
We next use \eq{RR} to express the Riemann tensor through the Ricci
and metric tensors and write down the product of two epsilons as
a determinant made of Kronecker deltas. Performing all convolutions
we obtain that \eq{std3} can be identically rewritten as

\beq
\left[g_{qs}\left(R\,\delta^i_p-2R^i_p\right)
-g_{ps}\left(R\,\delta^i_q-2R^i_q\right)\right]
m^pm^s\left(\nabla_{\!i}\,m\right)^q.
\la{std4}\eeq
Here the first term is zero because of \eq{covun1} while in the second
term we use $g_{ps}m^pm^s=1$. As a result we get

\beq
-\left(R\,\delta^i_q-2R^i_q\right)\left(\nabla_{\!i}\,m\right)^q
\la{std5}\eeq
which cancells exactly with \eq{ftd1}. Thus,
$\left(\nabla_{\!k}\right)^k_lV^l=0$, {\it q.e.d.}

\end{document}